\documentclass{iaus}
\usepackage{graphicx}

\title[$H_0$ from time delays]{The Hubble Constant from Gravitational
Lens Time Delays}

\author[Paul L. Schechter]%
{Paul L. Schechter$^1$%
  \thanks{Present address: Center for Space Research 37-664G,
77 Massachusetts Avenue, Cambridge, MA 02139, USA}
}

\affiliation{$^1$Department of Physics, Massachusetts Institute of
Technology, Cambridge, MA 02139, USA
email: schech@mit.edu
}

\pubyear{2004}
\volume{225}
\pagerange{1--8}
\date{?? and in revised form ??}
\jname{Impact of Gravitational Lensing on Cosmology}
\editors{Mellier, Y. \& Meylan,G. eds.}
\begin{document}

\maketitle

\begin{abstract}  Present day estimates of the Hubble constant
based on Cepheids and on the cosmic microwave background radiation are
uncertain by roughly 10\% (on the conservative assumption that the
universe may not be {\it perfectly} flat).  Gravitational lens time
delay measurements can produce estimates that are less uncertain, but
only if a variety of major difficulties are overcome.  These include a
paucity of constraints on the lensing potential, the degeneracies
associated with mass sheets and the central concentration of the
lensing galaxy, multiple lenses, microlensing by stars, and the small
variability amplitude typical of most quasars.  To date only one lens
meets all of these challenges.  Several suffer only from
the central concentration degeneracy, which may be lifted if one is
willing to assume that systems with time delays are either like better
constrained systems with non-variable sources, or alternatively, like
nearby galaxies.
\end{abstract}

\firstsection % if your document starts with a section,
\section{Context}

Were we to choose at random a hundred members of the IAU and ask each
of them to tell us the value of the Hubble constant and how it is
measured, few if any would refer to gravitational lens time delay
measurements.  Most would instead refer to observations of Cepheid
variables with HST or to observations of the CMB power spectrum with
WMAP.  These set the {\it de facto} standard against which time delay
estimates must be evaluated.

The Cepheids give a Hubble constant of $72$ km/s/Mpc with a 10\%
uncertainty \cite[(Freedman et al. 2001)]{2001ApJ...553...47F}, a
result that most of our randomly chosen astronomers would find relatively
straightforward.  But only a handful of them would be able to tell us
how the CMB power spectrum yields a measurement of the Hubble
constant.

In his opening talk, David {\sc Spergel} told us that three numbers are
determined to high accuracy by the CMB power spectrum: the height of
the first peak, the contrast between the heights of the even and odd
peaks, and the distance between peaks.  Ask anyone who calls
himself a cosmologist how many free parameters his world model has and
the number $N$ will be larger than three.  Spergel's
three numbers constrain combinations of those $N$ parameters, but do
not, in particular, tightly constrain the Hubble constant.  One must
supplement the CMB either with observations, or alternatively, with
non-observational constraints.

The effects of supplementing the observed CMB power spectrum with
other observations, and of adopting (or declining to adopt)
non-observational constraints, are shown in table 1.  All of the
numbers shown are taken from the analysis by
\cite{2004PhRvD..69j3501T}.  They use the first year data from the
Wilkinson Microwave Anisotropy Probe \cite[(Bennett et
al. 2003)]{2003ApJS..148....1B}, the second data release of the Sloan
Digital Sky Survey \cite[(Abazajian et al. 2004)]{2004AJ....128..502A},
the \cite[Tonry et al. (2003)]{2003ApJ...594....1T}, data for high
redshift Type Ia supernovae, and data from 6 other CMB experiments:
Boomerang, DASI, MAXIMA, VSA, CBI and ACBAR.

\begin{table}[t]\def~{\hphantom{0}}
  \begin{center}
  \caption{The CMB power spectrum and the Hubble constant}
  \label{tab:cmb}
  \begin{tabular}{llllll}\hline
     \multispan3{\hfil very-nearly flat \hfil} & 
     \multispan3{\hfil perfectly flat \hfil} \\\hline  
     observations & $~~~\Omega_{tot}$ & $100h$ & 
 ~~~~observations &  $\Omega_{tot}$ & $100h$ \\\hline
     WMAP1   & $1.086^{+0.057}_{-0.128}$ & $50^{+16}_{-13}$&
 ~~~~WMAP1   & $~1$                       & $74^{+18}_{-7}$ \\
     add SDSS2 & $1.058^{+0.039}_{-0.041}$ & $55^{+9}_{-6}$ & 
 ~~~~add SDSS2 & $~1$                       & $70^{+4}_{-3}$  \\
     add SNae & $1.054^{+0.048}_{-0.041}$ & $60^{+9}_{-6}$ &
 ~~~~add other CMB & $~1$                  & $69^{+3}_{-3}$  \\\hline
  \end{tabular}
 \end{center}
\end{table}

We see from the first line of table \ref{tab:cmb} that the value of
the Hubble constant derived from the WMAP1 data alone is a factor of
1.5 or 2.5 more uncertain than the Cepheid result, depending upon
whether the universe is taken to be perfectly flat or only very-nearly
flat.  Combining the galaxy power spectrum as measured with SDSS2
\cite[(Tegmark et al. 2004a)]{2004ApJ...606..702T} reduces the
uncertainties by factors of 3 and 2, respectively.  In the very-nearly
flat case, adding type Ia supernovae does little to change the
uncertainty but shifts the value of $H_0$ closer to the Cepheid value.
In the perfectly flat case, adding other CMB measurements reduces the
uncertainty in $H_0$ by another factor of 1.3.  Small deviations from
flatness produce substantial changes in the Hubble constant, with the
product $h\Omega_{tot}^5$ remaining roughly constant \cite[(Tegmark et
al. 2004b)]{2004PhRvD..69j3501T}.

Suppose we were to show table~\ref{tab:cmb} to our randomly chosen IAU
members and again ask them the value of the Hubble constant (and its
uncertainty).  Some would argue that the results for the very-nearly
flat case are so very-nearly flat that it is reasonable to adopt
perfect flatness as a working model.  We observe that $|\log
\Omega_{tot}| < 0.10$, when it might have been of order 100 (or
perhaps 3 if one admits anthropic reasoning).  But others would be
reluctant to take perfect flatness for granted.  The very sensitivity
of the Hubble constant to that assumption would be an argument against
adopting it.  The present author is, himself, ambivalent.

\section{$H_0$ from time delays {\it circa} 2003}

\begin{figure}
 \centerline{
 \scalebox{0.72}{
 \includegraphics{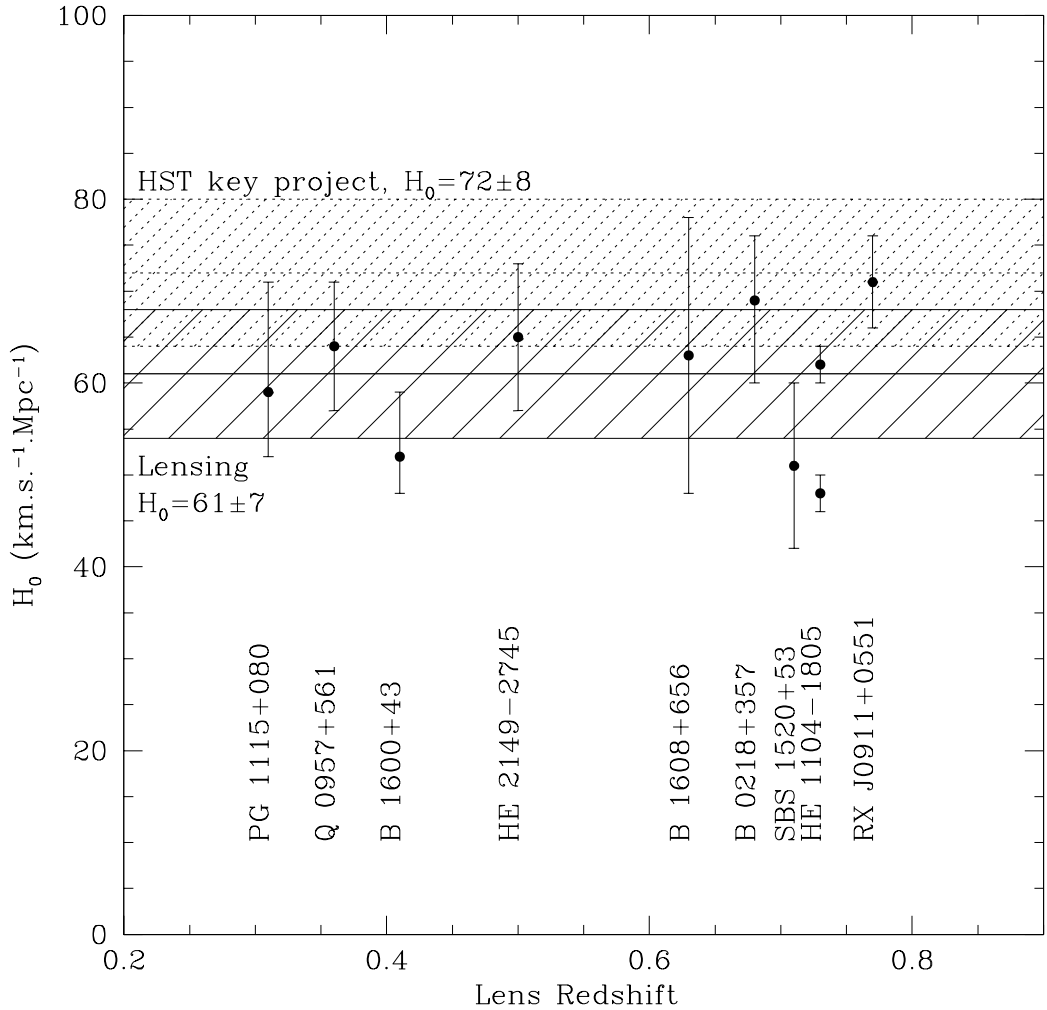}}
 }
  \caption{
  Time delay measurements of the Hubble constant, {\it circa} 2003
  \cite[(Courbin 2003)]{2003Courbin}.
  }\label{fig:courbin}
\end{figure}

\cite{1964MNRAS.128..307R} first proposed measuring the Hubble
constant using gravitationally lensed supernovae 15 years before the
discovery of the first gravitationally lensed quasar.  In the past two
years the status of Hubble constant measurements from time delays has
been reviewed by \cite{2003Courbin} and by \cite{2004mmu..symp..117K}.
Figure~\ref{fig:courbin}, taken from Courbin's review, summarizes his
report.  It should be noted that Courbin deliberately incorporated
results representing diverse approaches to computing $H_0$.
The average is one standard deviation smaller than the
Cepheid result with a comparable uncertainty.  It is also in agreement
with the CMB results of table~\ref{tab:cmb}.

But if we were to make such a figure today, we might find it less
reassuring.  The $H_0$ value for PG1115+080 in
figure~\ref{fig:courbin} is based on a model for the lensing galaxy
that is far from the fundamental plane for ellipticals.  The value for
RXJ0911+0554 is based on a faulty model \cite[(Schechter
2000)]{2009IAUS..201..000S} that the present author repudiates in
\S\ref{sec:multiple}.  And the values for HE1104-1805 are based on a
time delay that \cite{2003ApJ...594..101O} have shown to be too large
by a factor of two, driving the points off the top of the plot.  The
uncertainty in the Hubble constant, as derived from the scatter among
systems, would be considerably larger today than it is in
figure~\ref{fig:courbin}.

In what follows we discuss a number of major difficulties associated
with using time delay measurements to measure $H_0$.  All but one of
the systems in figure~\ref{fig:courbin} are subject to one or more of
these difficulties.  Before delving into the details, we review the
physics upon which the estimate of the Hubble constant is
based.\footnote{An excellent, although slightly outdated pedagogical
treatment of gravitational lensing can be found in
\cite{1999fsu..conf..360N}.}

\section{Lensing fundamentals}

\subsection{The 3 D's}
There is a sense in which time delay is the most fundamental
manifestation of gravitational lensing.  In the weak field limit a
gravitational potential produces an effective index of refraction that
increases the travel time for a photon.  Fermat's principle requires
that the photon travel along a path that is a minimum, a maximum or a
saddlepoint of the travel time.  The photon must detour around the
path it otherwise would have taken.  A balance is struck between
increased travel time associated with the detour (the geometric delay) and
decreased travel time due to the shallower
gravitational potential (the Shapiro delay),
\begin{equation}
{\rm {time \choose delay} = {geometric \choose delay} + {gravitational \choose delay} \quad .}
\end{equation}
If the size of the lens is small compared to the path length, we can
use the thin lens approximation.  The time delay can be expressed in
terms of a two-dimensional effective potential obtained by integrating the
gravitational potential $\Phi_{3D}$ along the line of sight,
\begin{equation}
\psi_{2D} = {D_{LS} \over D_S} \int_{observer}^{source} {2 \Phi_{3D}\over c^2 } {d \ell \over D_L} \quad. 
\end{equation}
It depends upon the angular diameter distances to the lens and to the
source, $D_L$ and $D_{S}$, and upon the distance from the lens to the
source, $D_{LS}$.  With this simplification all of gravitational lensing
boils down to taking derivatives of the time delay function,
\begin{equation}
\tau = {1+z_L \over H_0} {d_L d_S \over d_{LS}} \left[{1\over2}
(\vec \theta - \vec \beta)^2
- \psi_{2D}(\vec \theta)\right] \quad ,
\end{equation}
where $\vec\theta$ measures the position on the sky at which a ray
crosses the plane of the lens and $\vec\beta$ is the position on the 
sky of the source.  Here we use the dimensionless variants of the
angular diameter distances, $d_L$, $d_{S}$, and $d_{LS}$.  
The quantity $\vec \theta - \vec \beta$ is the amount by which
a ray is deflected. The geometric part of the time delay varies 
as the square of deflection in the limit of small deflections,
a straightforward consequence of the Pythagorean theorem.

Differentiating $\tau$ with respect to $\vec\theta$
and setting the gradient of $\tau$ to zero allows us to solve for
the minima, maxima and saddlepoints of the time delay.  It gives what is 
often call {\it the} lens equation,
\begin{equation}
\quad\vec \theta - \vec \beta - \vec \nabla \psi_{2D}  = 0 \quad .
\label{lens}
\end{equation}
The deflection, $\vec \theta - \vec \beta$, is equal to the gradient of
the two dimensional potential.

The magnification of an image is the ratio of the observed angular
size to the true angular size of an object.  We want something
like the ratio of $\Delta \vec \theta$ to $\Delta \vec \beta$, but
both of these are vectors.  Passing to derivatives, the quantity
$\partial \vec \theta / \partial \vec \beta$ gives a magnification
matrix that maps a vector at the position of the source, $\Delta \vec
\beta$, into a vector in the image plane, $\Delta \vec \theta$.
This matrix cannot be calculated from equation~\ref{lens} because the
two dimensional potential is a function of position in the lens plane,
not the source plane.  But its inverse can be calculated by taking
the derivative of equation~\ref{lens} with respect to $\vec \theta$,
giving the inverse magnification matrix,
\begin{equation}
{\partial \vec \beta \over \partial \vec \theta}  =
\Biggl(
{1 - {\partial^2 \psi \over \partial \theta_x^2}  \quad - 
{\partial^2 \psi \over \partial \theta_x \partial \theta_y}
\atop
- {\partial^2 \psi \over \partial \theta_x \partial \theta_y} 
\quad 1 - {\partial^2 \psi \over \partial \theta_y^2}}
\Biggr) \quad .
\end{equation}
This is readily inverted to give the magnification matrix, $\partial
\vec \theta / \partial \vec \beta$.  Since the magnification matrix is
symmetric, there is a choice of coordinates for which it is diagonal.
Those diagonal elements are in general unequal, stretching or
compressing the image more in one direction than the other.  The
magnification matrix might equally well be called the distortion
matrix.

It is often the case that the source and its images are very much
smaller than the resolution of our observations.  Gravitational
lensing conserves surface brightness, but as the solid angle subtended
by the image is usually different from that subtended by the source,
the observed flux is also different.  The scalar magnification
factor, $\mu$, is given by
\begin{equation}
\mu = \Bigl|{\partial \vec \theta \over \partial \vec \beta}\Bigr|   \quad .
\end{equation}

The observable consequences of lensing are mnemonized with three
``D's,'' each corresponding to a different derivative of the time
delay function.  The first derivative of the delay function gives a
deflection.  The second derivative gives a distortion.  And the
``zeroth'' derivative is just the delay function itself.

\subsection{Modeling the lens}

\begin{table}[b]\def~{\hphantom{0}}
  \begin{center}
  \caption{Constraints}
  \label{tab:constraints}
  \begin{tabular}{ll}\hline
Delays: & $1\times[\# images - 2 ]$ \\
Deflections: & $2\times[\# images - 1]$ \\
Distortions: & $3\times[\# images - 1]$ \\
\multispan2 {\hfil (or for unresolved sources) \hfil} \\
magnifications: & $1 \times [\#images - 1]$ \\ \hline
  \end{tabular}
 \end{center}
\end{table}

The previous subsection illustrates the power of mathematics:
an enormous amount of interesting astrophysics can be swept under the
rug of a single function, the two dimensional gravitational potential,
$\psi(\vec\theta)$.  If we knew $\psi$ {\it a priori} we would long
since have measured the Hubble constant.  Instead we must construct
models for our potentials, adjusting them to fit whatever observations
can be brought to bear.  \cite{2000AJ....119..439W}  have developed
a method for modeling lenses that avoids making detailed assumptions
about the shape of the lensing potential (see also {\sc saha
\& williams'} contribution to the present proceedings), at the price of
non-uniqueness.  The present author takes the view that we know
something about galaxy potentials and therefore would do well to start
by assuming lensing galaxies are much like nearby galaxies.

A model which is both very simple and very useful is the singular
isothermal sphere in the presence of an external tide,
\begin{equation}
\psi_{2D}(\vec\theta) = br + {\gamma \over 2}r^2 \cos 2(\phi - \phi_\gamma)
\quad .
\end{equation}
The leading (monopole) term is the projection of the 
isothermal's potential.  In our case the strength of the isothermal is
parameterized by $b$, the radius of its Einstein ring.  The tidal term
is a quadrupole, and is parameterized by a dimensionless strength,
$\gamma$, and an orientation $\phi_\gamma$.  Note that here we
represent angular position on the sky, $\vec\theta,$ in terms of polar
coordinates $r$ and $\phi$, with $r$ measured in radians.

The singular isothermal sphere gives the flat rotation curves
characteristic of nearby galaxies.  A substantial fraction of
quadruple gravitational lenses are the result of strong tides from
a group or cluster of galaxies of which the lens is a member.  In the
absence of these tides the lenses would produce only double images.
The intrinsic flattening of the lensing galaxy also produces a
quadrupole term (which has a different radial dependence), but the
tidal contribution is usually considerably larger.

Our simple model does great injustice to the complexity of individual
lenses, but complexity is a luxury we cannot afford.  A parameterized
model can have at most as many parameters as the available observable
quantities and we are quite limited.  The numbers of observables are
summarized in table~\ref{tab:constraints}.  Delay is a scalar, deflection
is a two dimensional vector and the symmetric distortion matrix has
three independent elements.  We have subtracted one deflection from
our available constraints because we do not know the position of the
source.  Likewise we subtract one distortion because we do not know
the size and shape of our source (taking it to be elliptical).  By the
same argument we subtract one delay because we do not know the
absolute time at which some event used to calculate delays actually occurred.
In the case of delays we subtract a second delay because we need it to
calculate the Hubble constant.  It cannot be used {\it both} to constrain
our model {\it and} to give $H_0$.

The situation is even worse if the images are not resolved by our
telescopes.  In such cases we measure only the fluxes of the images
and get only one number from each distortion rather than
three.  Consider the case of a lensed quasar with two unresolved
images.  After discounting for the unknown properties of the source
and reserving one delay for measuring $H_0$ we have only three
constraints for our model.  Using even our very simple model we are
guaranteed a perfect fit, with no additional degrees of freedom
left to check goodness of fit.

With no sanity check on our model, such doubles can be used only with
extreme caution.  Something might be amiss and we would
have no way of knowing it.  Moreover we cannot add additional parameters
(allowing, for example, for the shape of the galaxy) because we have
exhausted our supply of constraints.  Four of the nine systems in
figure~\ref{fig:courbin} are just such doubles: B1600+43, HE2149-2745,
SBS1520+53 and HE1104-1805.

The situation is slightly improved for four image systems.  It is
likewise slightly improved if there are multiple sources, as is
sometimes the case for radio AGN.  In constructing a list of major
difficulties faced in measuring $H_0$, the first would be the
paucity of constraints.

\begin{figure}[t]
 \centerline{
 \scalebox{0.72}{
 \includegraphics{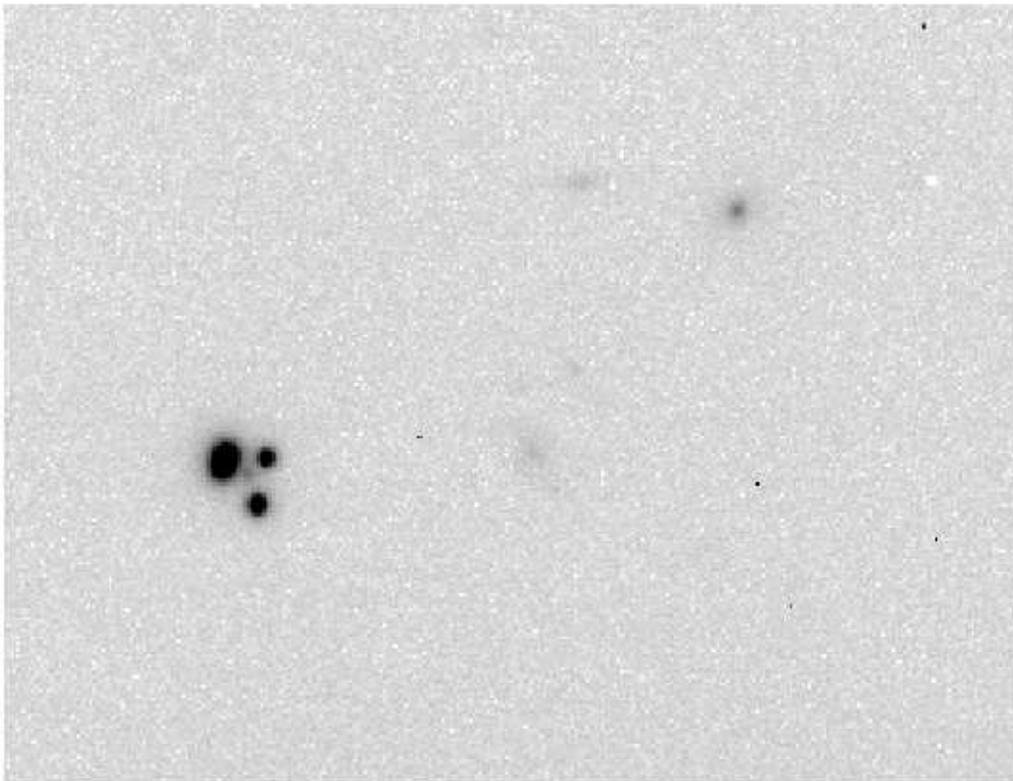}}
 }
  \caption{
  The quadruple lens PG1115+080, observed with the Baade 6.5-m
  telescope.  The small group of galaxies to the upper right lies
  very close to the position predicted from the positions of
  the four quasar images.  It lies at the same redshift as the
  lensing galaxy.
  }\label{fig:pg1115}
\end{figure}

\section{Problem lenses}

\subsection{Kochanek's time delay formulation}

The system PG1115+080 illustrates another of the major difficulties in
time delay measurements of $H_0$.  The mathematics is simpler if
we diagonalize the magnification matrix and introduce two dimensionless
quantities, the convergence $\kappa$ and the shear $\gamma$.
\begin{equation}
{\partial \vec \theta \over \partial \vec \beta}  =
\Biggl({
{1 \over 1 - \kappa - \gamma}  \quad 0 \quad
\atop
\quad 0 \quad {1 \over 1 - \kappa + \gamma}
}\Biggr) \quad .
\end{equation}
The convergence is the two-dimensional Laplacian of the
two-dimensional potential,
\begin{equation}
\kappa = {1 \over 2} \nabla^2_\theta \psi \quad ,
\end{equation}
and is therefore proportional to the surface mass density.

For zero shear, a convergence of unity gives infinite
magnification in both directions.  For zero convergence (which means
zero surface density), a shear of unity gives infinite stretching in
one direction.  For a point mass there is unit shear everywhere on its
Einstein ring.

\cite[{\sc Kochanek} (2002; present proceedings)]{2002ApJ...578...25K}
has shown that with these definitions we can write an expression for
the difference in time delay between two images in a circularly
symmetric system that depends to first order only upon the positions
of the images with respect to the lens and the convergence averaged
over the annulus bracketted by the two images.
\begin{equation}
\tau_B - \tau_A  
\simeq \left[ {1+z_l   \over H_0    } \right] 
       \left[ {d_l d_s \over d_{ls} } \right] 
(|\vec\theta_A|^2 - |\vec\theta_B|^2)
  \times 
\left[ 1- \left< \kappa \right> + {\cal O} \left( 
{|\vec\theta_B| - |\vec\theta_A| \over |\vec\theta_B| + |\vec\theta_A| } 
\right)^2 \right] \quad .
\label{taukappa}
\end{equation}
For power law density profiles, $\rho \sim r^{-\eta}$, we have
\begin{equation}
\left< \kappa \right> \simeq {3 - \eta \over 2} \quad .
\label{rhoeta}
\end{equation}
Notice that if $\eta = 2$, as would hold for an isothermal sphere, the
average value of the convergence is one half.  As $\eta \rightarrow
3$, approaching the central concentration for a point mass, $\left<
\kappa \right> \rightarrow 0$ and the predicted time delay is twice as
long as in the isothermal case.

\subsection{The mass sheet degeneracy}

Kochanek's parameterization illuminates two different difficulties in
measuring $H_0$ from time delays.  Figure~\ref{fig:pg1115} shows
PG1115+080 and its immediate surroundings.  A fit of our simple model
to the image positions indicates that the source of the tide lies
along a line passing through the lensing galaxy and the upper right
corner of the figure.  At the cost of one additional parameter, we can
substitute a second isothermal sphere for the tidal shear.  This very
much improves the fit, and the derived position is coincident with the
small group of galaxies to the upper right of the lensed system.  The
inferred velocity dispersion for the second isothermal sphere is 375
km/s \cite[(Schechter et al. 1997)]{1997ApJ...475L..85S}, typical of a
small group.  The observation of a group at the position ``predicted''
by the simple model inspires confidence in it.

\begin{table}[t]\def~{\hphantom{0}}
  \begin{center}
  \caption{Exponents for power law density profiles: $\rho \sim r^{-\eta}$}
  \label{tab:exponents}
  \begin{tabular}{lll}\hline
lensing galaxy  & ~~$\eta$              & reference \\\hline
JVAS0218+357  & $1.96^{+0.02}_{-0.02}$& \cite{2004MNRAS.349...14W} \\
Q0957+561  & $1.84$                & \cite{1999ApJ...520..479B} \\ 
MG1131+0456 & $2.40^{+0.2}_{-0.2}$  & \cite{1995ApJ...447...62C} \\ 
PMN1632-0033 & $1.91^{+0.02}_{-0.02}$& \cite{2004Natur.427..613W} \\ 
MG1654+1346 & $1.90^{+0.16}_{-0.01}$&  \cite{1995ApJ...445..559K} \\ 
CLASS1933+503  & $1.86^{+0.17}_{-0.11}$& \cite{2001ApJ...554.1216C} \\\hline
\hfil $< \quad >$ & 1.98 & \\
  \end{tabular}
 \end{center}
\end{table}

But if the group is isothermal, it has a convergence associated with
it -- in this case approximately 10\%.  This decreases the predicted
time delay by 10\%.  Small numbers make measurement of the actual
velocity dispersion difficult, but it is consistent \cite[(Tonry
1998)]{1998AJ....115....1T} with the tidal estimate.  Many of the
quadruple lenses live in groups and clusters with convergences at the
position of the lens of 10, 20 and even 30\%.  The corrections are 
substantial, and we must wonder whether we have got the cluster
convergence correct.  Two of the systems in figure~\ref{fig:courbin}, Q0957+561
and RXJ0911+0554, have inferred convergences in the 20 - 30\% range.  This is
a manifestation of a difficulty more generally known as the mass sheet
degeneracy (\cite[Gorenstein, Shapiro, \& Falco
1988]{1988ApJ...327..693G}; \cite[Saha 2000]{2000AJ....120.1654S}).
There is no way of knowing (from deflections and distortions) whether
or not a mass sheet is present, but it affects the predicted time
delays by changing the mean convergence.

\subsection{The central concentration degeneracy}

Equations \ref{taukappa}  and \ref{rhoeta} tell us that
the interpretation of the time delay for PG1115+080 depends crucially
upon the internal structure of the galaxy.  An error in the radial
exponent of plus or minus 0.1 produces a 10\% change in the predicted
time delay as compared to an isothermal model.  
Fortunately we can test the isothermality hypothesis.
For an isothermal sphere, the angular radius of
the Einstein ring, $b$, is a directly proportional to the
square of the velocity dispersion, $\sigma^2$:
\begin{equation}
b = {d_{LS} \over d_S} 4 \pi {\sigma^2 \over c^2}
\label{impliedsigma}
\end{equation}
With an Einstein ring radius $b = 1.\!\!^{\prime\prime}15$, a
lens redshift $z_L = 0.31$ and a source redshift $z_S = 1.71$ we
predict $\sigma_{SIS} = 232 $ km/s.  This must be reduced by a factor
$\sqrt{(1-\kappa_{group}})$ which amounts to a 5\% correction in the
present case.  We can test our prediction, but the measurement is a
difficult one.  The lensing galaxy is crowded by four very much
brighter images.  \cite{1998AJ....115....1T} measures a velocity
dispersion of $281 \pm 25$ km/s, very much larger than predicted.
\cite{2002MNRAS.337L...6T} use Tonry's measurement in modeling PG1115+080
and derive a value for $H_0$ very much larger than under the
isothermal hypothesis.  Their value is the one plotted in
figure~\ref{fig:courbin}.

\begin{figure}[t]
 \centerline{
 \scalebox{0.72}{
 \includegraphics{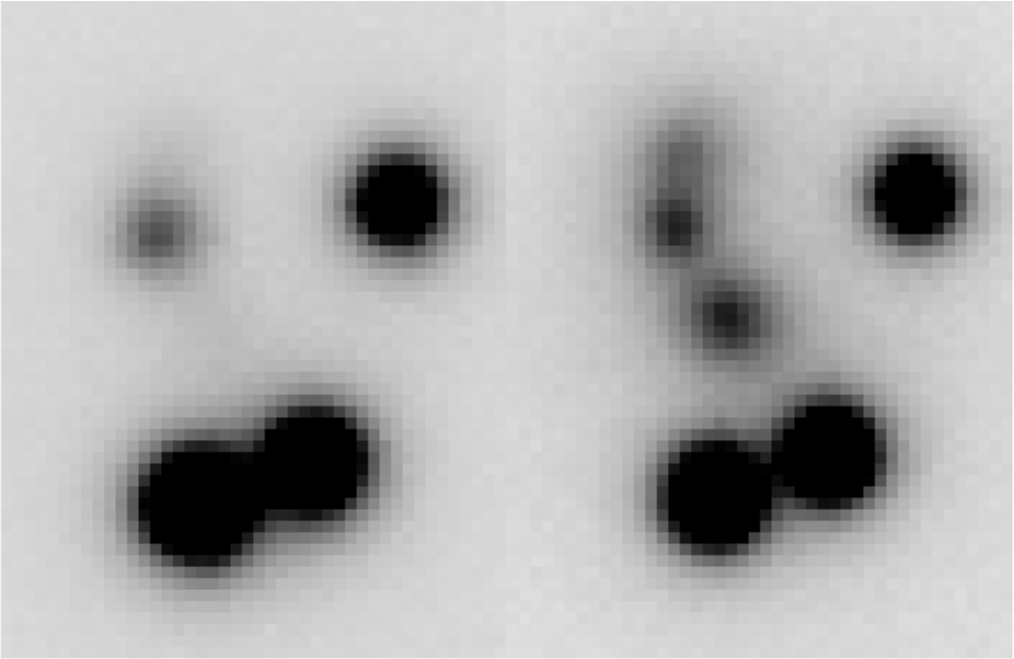}}
 }
  \caption{
  The quadruple system HE0230-2130, observed with the Baade 6.5-m telescope
  through $g'$ (left) and $i'$ (right) filters.  The two red objects
  are lensing galaxies.  The image between the two galaxies is a
  saddlepoint of the time delay function.
  }\label{fig:he0230}
\end{figure}

Why are we so reluctant to abandon the isothermal hypothesis for
PG1115+080?  First, because velocity dispersion estimates from
equation \ref{impliedsigma} for an ensemble of lensing galaxies are
consistent with the fundamental plane relation for non-lensing
ellipticals.  PG1115+080 is in no way unusual \cite[(Kochanek et
al. 2000)]{2000ApJ...543..131K}, but would be if we adopted the direct
measurement.

A second argument is that the lenses for which we {\it can} measure
the radial exponent are very nearly isothermal.  Lenses with multiple
sources, rings or central images all break the central concentration
degeneracy and permit measurement of the radial exponent $\eta$.  The
results for six systems are shown in table~\ref{tab:exponents} (see
also {\sc Wayth's} contribution to the present proceedings).  While
there are differences in the way $\eta$ was calculated for each of
these systems, the results are consistent with isothermal.
Unfortunately only one of these systems, JVAS0218+357, is also a
system that has a measured time delay.

A third argument is that a great many nearby galaxies have been
studied, and their potentials are consistent with isothermals.  This
is nicely shown in a figure published by \cite{1999ApJ...516...18R}.
For a sample of twenty bright ellipticals, the circular velocity
inferred from the velocity dispersion declines only slightly over a
factor of 30 in radius.  The corresponding decline in PG1115+080,
computed from its measured central dispersion and the Einstein ring
radius, is very much larger.  It is nonetheless true that the
potentials for nearby ellipticals are only very-nearly isothermal and
not perfectly isothermal.  The central concentration degeneracy
qualifies as another major difficulty associated with time delay
estimates of $H_0$.

\subsection{Multiple lenses}\label{sec:multiple}

Another problem associated with modeling lensing galaxies is
illustrated by the system HE0230-2130.  There is not one lensing
galaxy, but two, as seen in figure~\ref{fig:he0230}.  The second,
fainter galaxy provides the quadrupole moment that makes this a
4-image system and not a 2-image system.  When we see two galaxies
separated by only a few kpc, we must wonder how the non-baryonic
matter is distributed.  Are the halos for the two galaxies distinct?
Or have they merged?  How many components should we model?  And where
should their centers be?  In such a case something with elements of
the \cite{2000AJ....119..439W} form-free approach might be preferable
to a straightforward parameterized model.

\begin{figure}[t]
 \centerline{
 \scalebox{0.72}{
 \includegraphics{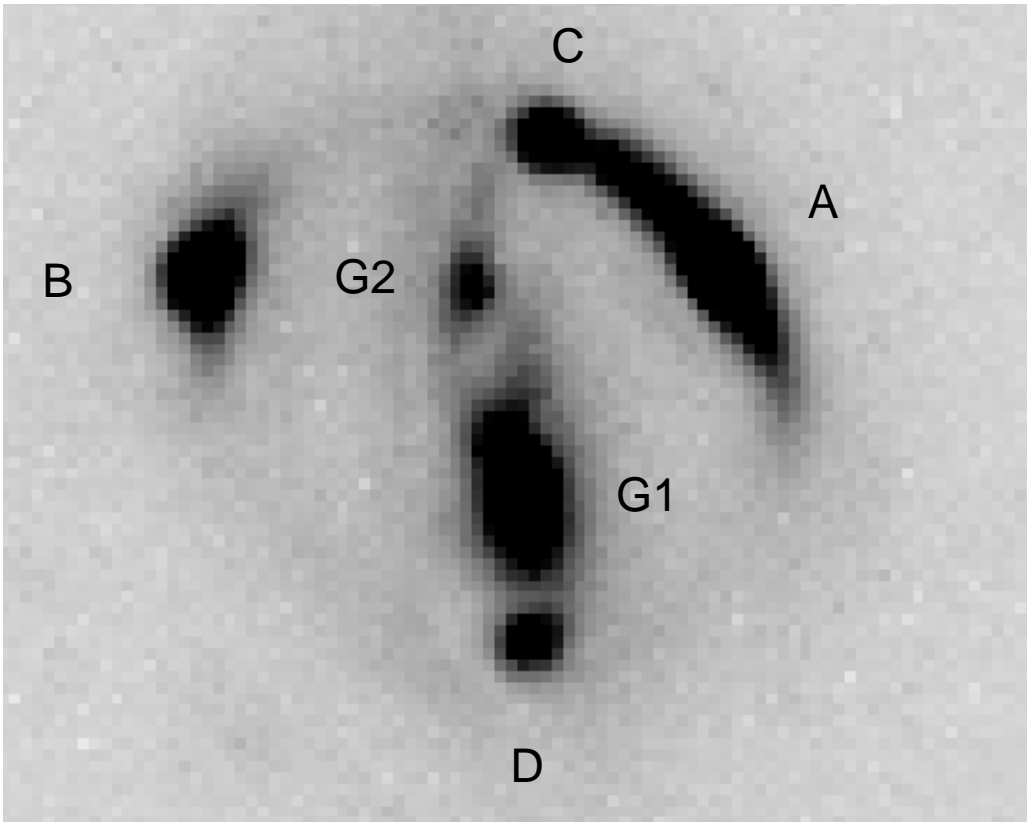}}
 }
  \caption{
  The quadruple system CLASS1608+656, observed with the Hubble Space
  Telescope 
  through the $F814W$ filter.  The lensing galaxies are marked
  $G1$ and $G2.$  A dust lane can be seen encircling them.
  }\label{fig:q1608}
\end{figure}

The system with the best time delays is CLASS1608+656.
\cite{2002ApJ...581..823F} have measured four beautiful radio
lightcurves.  All four time series faithfully reproduce the many bumps
and wiggles.  The multiple delays in this case are good enough to
provide meaningful additional constraints to the deflections and
distortions.  But modeling the system is not straightforward.  As
with HE0230-2130, there
are two lensing galaxies (figure~\ref{fig:q1608}) that appear to be
interacting.  A dust lane encircles them both.  Again the question of
how the dark matter might be distributed looms as crucial.
\cite{2003ApJ...599...70K} have worked exceedingly hard to constrain
this system, measuring positions for the images, the shape of the
ring, and measuring the velocity dispersion of the more massive lens.
We can nonetheless imagine a devil's advocate coming up with a
plausible dark matter distribution uncorrelated with the observed
galaxies that gives a very different value for the Hubble constant.

\begin{figure}[t]
 \centerline{
 \scalebox{0.72}{
 \includegraphics{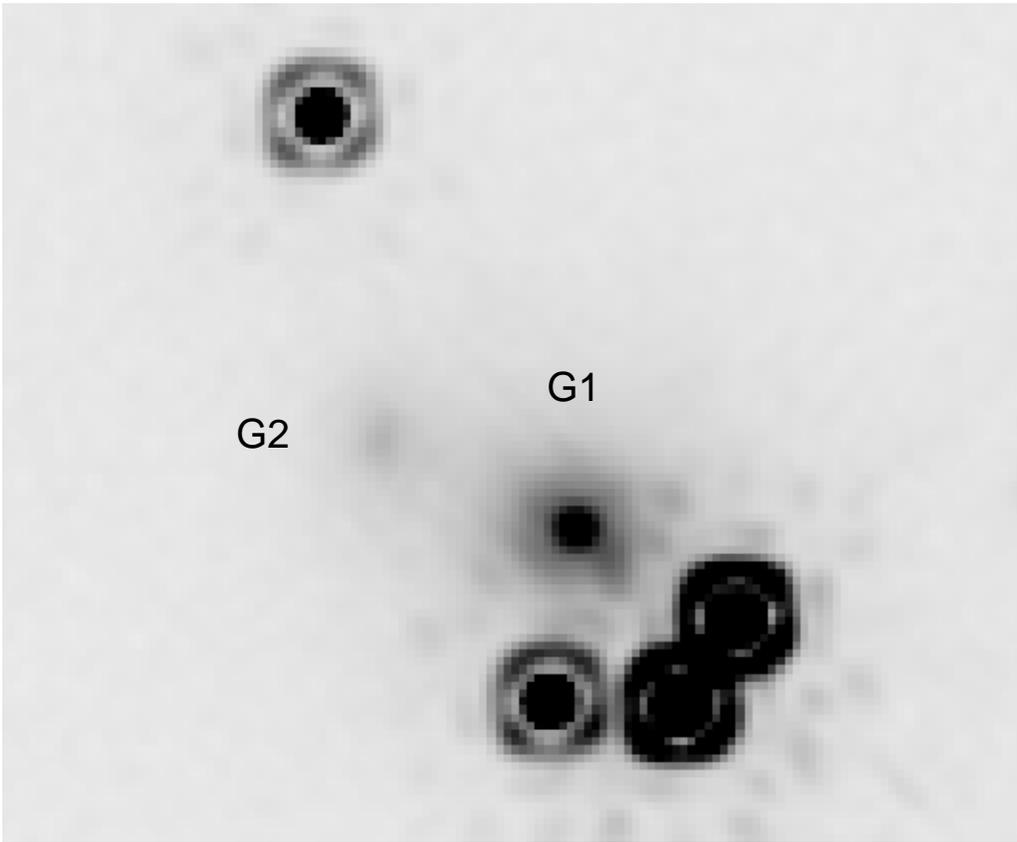}}
 }
  \caption{
  The quadruple system RXJ0911+0554, observed by the
  CASTLES consortium with the Hubble Space
  Telescope through the $F160W$ filter. The primary lensing galaxy $G1$
  has a dwarf satellite $G2$.  Allowing for the mass of the dwarf 
  changes the derived value of $H_0$ by 10\%.
  }\label{fig:rxj0911}
\end{figure}

The problem of multiple lenses can be serious even when one of the
galaxies is very much smaller than the principal lens.  Consider the case
of RXJ0911+0554, shown in figure~\ref{fig:rxj0911},  where the primary
lensing galaxy has a faint companion.  Given its faintness and our
aversion to adding additional parameters, we might be tempted to
ignore it, as did \cite{2009IAUS..201..000S} in the model used by
Courbin to construct figure~\ref{fig:courbin}.  But allowing for a
mass at the position of this dwarf companion changes the predicted
time delay by 10\%.  Though the smaller galaxy is a factor of 10
fainter than the primary lensing galaxy, its effect is to move the
center of mass closer to the midpoint of the images, decreasing the
differences in path length.

There is an irony here in that the multiple lenses work to our
advantage in producing (or adding to) the quadrupole moments that give
us 4 images rather than 2.  But at the same time they make modeling
considerably more difficult.

\subsection{Microlensing}

\begin{figure}
 \centerline{
 \scalebox{0.72}{
 \includegraphics{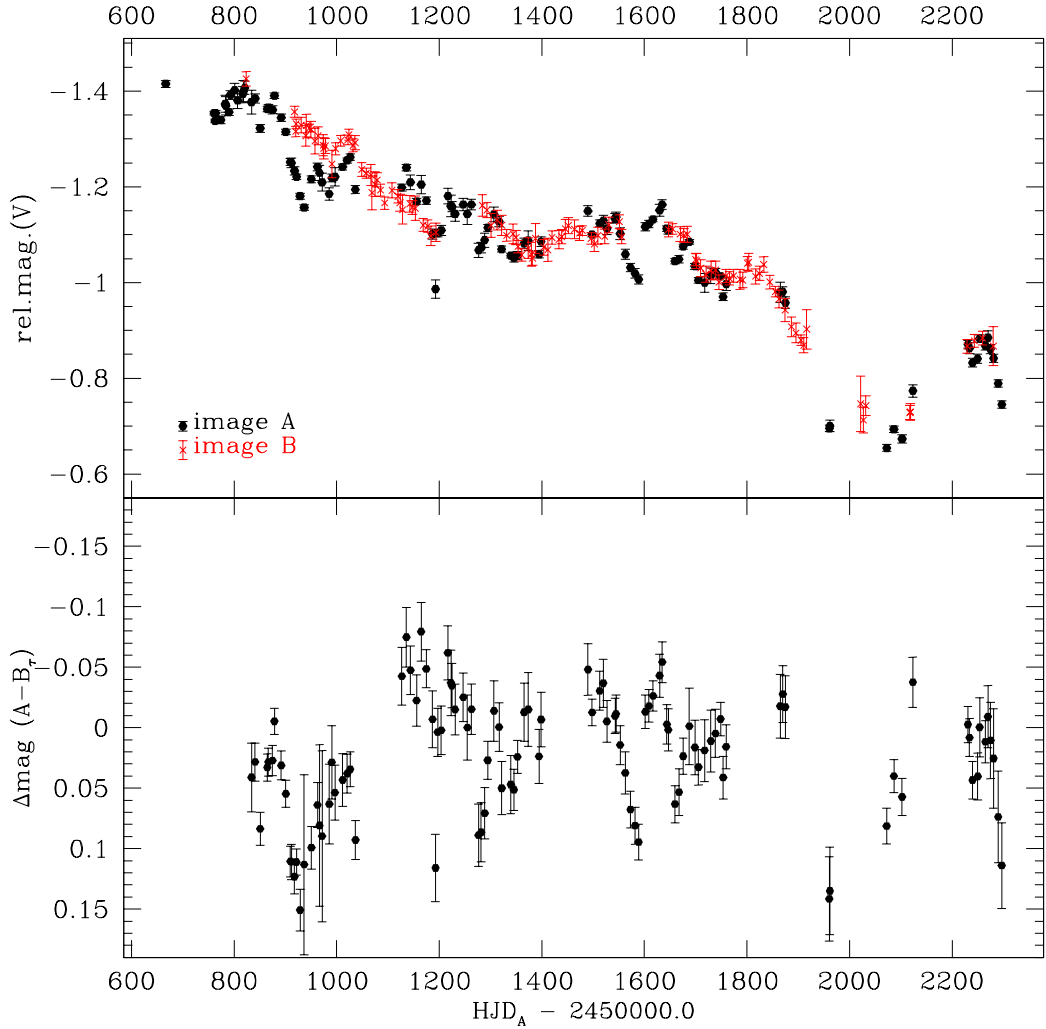}}
 }
  \caption{ Lightcurves for the two components of HE1104-1805, taken
  with the OGLE 1.3-m telescope \cite[(Wyrzykowski et
  al.\ 2003)]{2003AcA....53..229W}.  Note that while the measurement
  uncertainties are larger for the fainter, $B$ image, the scatter is
  larger for the brighter, $A$ image.  }\label{fig:he1104}
\end{figure}

In his contribution to the present proceedings, Lutz {\sc Wisotzki}
presents spectroscopic evidence for microlensing of quasars.  This
has important consequences (unfortunately negative)
for time delay estimates of $H_0$.  It means that there are
uncorrelated variations in the fluxes of images, especially in the
optical continuum.  The timescale for these fluctuations depends upon
the size of the Einstein ring of the microlenses and upon the relative
velocity of the quasar and the lensing galaxy.  Taking the lenses to
be a solar mass and the relevant velocities to be of order 300 km/s
gives timescales of order 10 - 30 years.  This ought not to
matter for quasars lensed by galaxies, whose time delays are at most
one year.  But there have been repeated instances of uncorrelated
variations in quasar lightcurves on considerably shorter timescales
(e.g. \cite[Burud et al. 2000]{2000ApJ...544..117B}), in some cases
causing considerable dispute about which points ought and ought not to
be included in a time delay measurement and about the nature of the
fluctuations.

Seven years ago the present author prevailed upon Andrzej Udalski to
monitor HE1104-1805 for a time delay with the OGLE telescope
\cite[(Udalski, Kubiak, \& Szyma{\'n}ski 1997)]{1997AcA....47..319U}.  The
OGLE group was at that time focussing exclusively on LMC and galactic
bulge microlensing, but HE1104-1805 is at a very different right
ascension.  Three years of superb OGLE data were obtained in which no
believable correlated variations were observed.  But there were
appreciable uncorrelated variations, most of which were attributable
to the brighter of the two images.  After three years we admitted
defeat and wrote a paper about microlensing \cite[(Schechter et
al. 2003)]{2003ApJ...584..657S} which was, after all, OGLE's primary
mission.  The OGLE telescope was out of commission for upgrades during
much of the following year; a few data points were obtained but were
never reduced.  But by then \cite{2003ApJ...594..101O} had begun
monitoring HE1104-1805 at the Wise Observatory.  They used our data in
conjunction with theirs to give a time delay that differed by a factor
of two from the one used in figure \ref{fig:courbin}.  After Ofek and
Maoz published their delay, \cite{2003AcA....53..229W} reduced the
remaining OGLE data, confirming their value and giving the lightcurves
in figure~\ref{fig:he1104}.

Notice that the error bars are smaller for the $A$ image, yet the
lightcurve is smoother for the $B$ image.  Image $A$ appears to vary
with an amplitude of 0.06 magnitudes on a timescale of one week.  The
interpretation of these fluctuations is somewhat speculative, but
there is no question that microlensing represents an another
difficulty for time delay measurements of $H_0$.

\subsection{The myth of quasar variability}

One last difficulty associated with estimating $H_0$ is
implicit in the fact that only 10\% of the hundred or so known lensed
quasars appear in figure~\ref{fig:courbin}.  When we think of quasar
variability we think of 3C273 or 3C279 and variations of a
magnitude or more.  But those quasars are well known precisely
because of their variability.  Most quasars are considerably more
boring.  Systematic studies of quasar variability at optical
wavelengths consistently give rms fluctuations of order 10 - 15\% on
proper timescales of one year (\cite[Cristiani et
al. 1996]{1996A&A...306..395C}; \cite[Vanden Berk et
al. 2004]{2004ApJ...601..692V}).

The good news is that a few quasars vary more than this.  The bad news
is that most vary even less than this.  The statistics may be somewhat
better for radio quasars, but then radio loud systems constitute only
10\% of all quasars.  While the small amplitude of quasar variations
may give graduate students (mostly those working at radio wavelengths)
an opportunity to show their skill in beating down their observational
errors, flat lightcurves don't lead to offers of prestigious postdocs.
It should be noted that there is considerable room for improvement in
the optical lightcurves, for which the observational errors are much
larger than photon statistics would imply.  Our list of
major difficulties in measuring $H_0$ has grown to include the following:
\begin{itemize}  
\item the paucity of constraints,
\item the mass sheet degeneracy,
\item the central concentration degeneracy,
\item multiple lenses,
\item microlensing and
\item small variability amplitudes.
\end{itemize}

\section{A golden lens}

The difficulties we have described are sufficiently common that almost
every lensed quasar is subject to at least one of them.  In general
(though there are important exceptions) the double systems suffer from
a paucity of constraints.  Many quadruples suffer either from the  mass
sheet degeneracy (due to a nearby cluster) or from multiple lensing
galaxies.  Only a few systems (mostly radio loud quasars) have
multiple sources permitting unambiguous determination of a radial
density exponent.  Are there {\it any} lenses which are not
obviously subject to one or another of these difficulties?

A decade ago we imagined finding a ``perfect'' \cite[(Press 1996)]
{1996IAUS..173..407P} or ``golden'' lens \cite[(Williams \& Schechter
1997)]{1997A&G....38e..10W} that would permit determination of $H_0$
with small uncertainty.  A lens that was not subject to any of the
above difficulties might reasonably qualify for certification as a
golden lens.

\begin{figure}
 \centerline{
 \scalebox{0.72}{
 \includegraphics[angle=270]{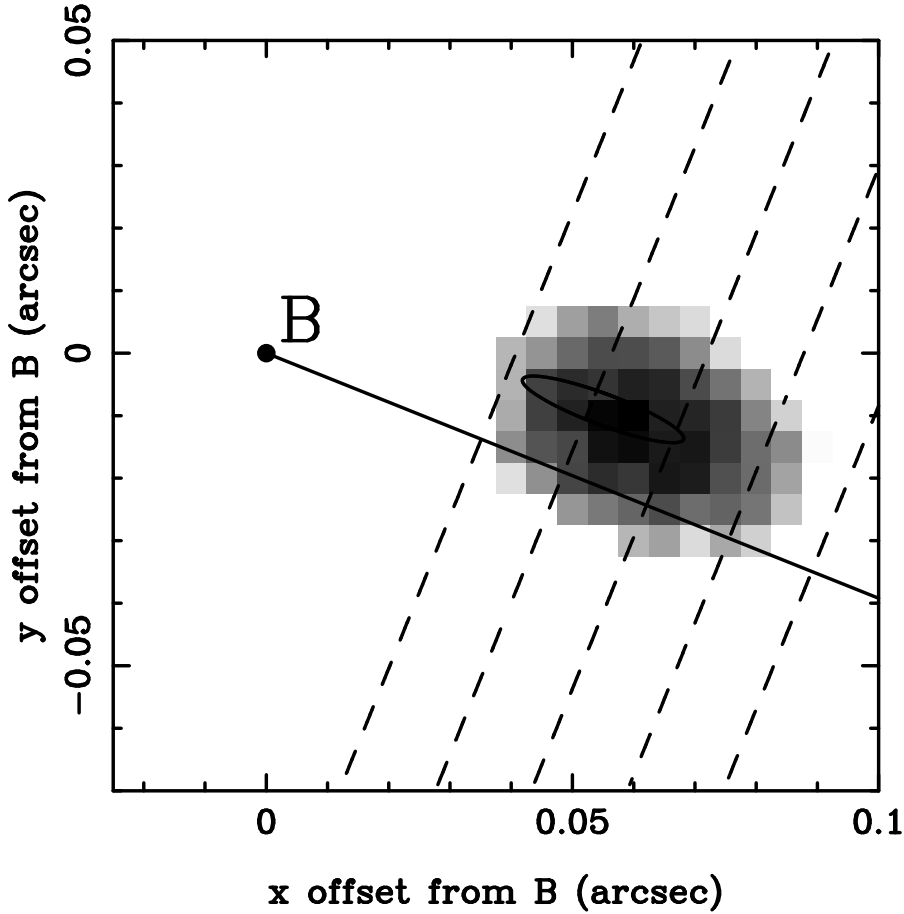}}
 }
  \caption{The shading of the pixels indicates relative likelihood for
  the optical position of the galaxy lensing JVAS0218+357 \cite[(York
  et al. 2004)]{2004York}.  The solid line is drawn from the $B$ image
  toward the $A$ image, which lies beyond the plot boundary.  The
  diagonal dashes indicate loci of constant Hubble constant, starting
  with $H_0 = 90$ km/s/Mpc on the left and decreasing in steps of $10$
  km/s/Mpc.  The elliptical contour gives the $2\sigma$ confidence
  region for the center 
  of the lensing potential, as derived from radio observations of the
  lensed radio source by \cite{2004MNRAS.349...14W}.\label{fig:york}
  }
\end{figure}

The system JVAS0218+357 \cite[(Patnaik et
al. 1992)]{1992LNP...406..140P} is such a system.  It is a radio
double that, at VLBI resolution, has a core and a blob \cite[(Biggs et
al. 2003)]{2003MNRAS.338..599B} of the sort that radio astronomers
fancifully call jets.  The double source breaks the concentration
degeneracy, and implies an isothermal
potential. \cite{2004MNRAS.349...14W} have modeled the radio data,
finding the radial exponent and the center of symmetry of the lensing
potential (see also {\sc Wucknitz'} contribution to the present
proceedings).  The time delay has been measured with better than 5\%
accuracy \cite[(Biggs et al. 1999)]{1999MNRAS.304..349B}.

But even this system is not 24 carat gold. The pairs of images are
separated by only $0.\!\!^{\prime\prime}33$, implying a low luminosity
lens, almost certainly not an elliptical.  \cite{2004York} have
co-added data from a large number of HST orbits (see also {\sc
Jackson and York's} contribution to the present proceedings).  After careful
PSF subtraction of the two quasar images, their data clearly shows an
M101-like spiral.  The residuals from PSF subtraction crowd the
nucleus of the galaxy, making it difficult to measure its centroid
accurately.  The position of the center is crucial because of the
strong dependence of the differential delay on the image distances.
The agreement between the York et al. optical position and the
Wucknitz center of symmetry is excellent (figure~\ref{fig:york}).  The
radio position gives $H_0 = 78 \pm 6$ km/s/Mpc ($2\sigma$
uncertainty).

The work on JVAS0218+357 is remarkable for the range of astronomical
techniques and resources that have been brought to bear on the system.
It was discovered \cite[(Patnaik et al. 1992)]{1992LNP...406..140P}
and monitored \cite[(Biggs et al. 1999)]{1999MNRAS.304..349B} at radio
wavelengths using the VLA and Merlin.  The global VLBI data
\cite[(Biggs et al. 2003)]{2003MNRAS.338..599B} are crucial to the
modeling.  The redshift for the lens \cite[(Browne, Patnaik,
Walsh, \& Wilkinson 1993)]{1993MNRAS.263L..32B} required large
ground based optical telescopes.  Measuring the position of the lensing
galaxy required HST \cite[(York et al. 2004)]{2004York}, as is the
case for most lenses.

\section{Summary and Prospects}

By the standards set forth in this paper, only one gravitationally
lensed quasar qualifies as a golden lens.  Taken by itself it gives
a Hubble constant with a formal uncertainty as small or smaller than
that obtained with Cepheids.  But prudence demands at least one more
golden lens before declaring that lenses do as well as (let alone better
than) Cepheids.

The remaining systems with time delay measurements are not without
value.  The central concentration degeneracy may reasonably be lifted
by appeal to other (non-variable) lensed systems for which the central
concentration {\it can} be measured, or alternatively, by appeal to
measurements of gravitational potentials of nearby elliptical
galaxies.  The present author would guess that the systematic errors
associated with either of these assumptions would introduce an error
of perhaps 5\% in the Hubble constant.  This would rehabilitate those
optical quadruple systems that suffer only from the central
concentration degeneracy.

Optical double systems suffer from both the central concentration
degeneracy and the paucity of constraints.  As they have no
internal redundancy, their usefulness will depend upon the scatter in
$H_0$ values observed from a good-sized sample of such systems.

In the case of B1608+656, the outstanding question is whether the data
are so good that, merging lenses notwithstanding, all plausible
models give the same value of $H_0$.  What is needed now is the effort
of a ``loyal opposition'' to search the far corners of model space.

New telescopes and upgrades to existing telescopes are certain to
produce new gravitational lenses, and with them we will eventually see
a Hubble constant that is less uncertain than that obtained from
Cepheids.  Even those who take it as a matter of faith that the
universe is perfectly flat may find such a direct measurement
of $H_0$ interesting.

\begin{acknowledgments}
The author gratefully acknowledges support from the US National
Science Foundation under grant AST02-06010, and thanks the members
of the organizing
committee for their good efforts.
\end{acknowledgments}

\end{document}